\documentclass[aps,prl,preprint,footinbib]{revtex4-1}

\usepackage{graphicx}
\usepackage{hyperref}
\usepackage{amsmath}

\renewcommand{\figurename}{Figure}

\begin{document}


\title{Self-learning analytical interatomic potential describing laser-excited silicon}

\renewcommand{\figurename}{FIG.}



\author{Bernd Bauerhenne}
\email{bauerhenne@uni-kassel.de}
\author{Vladimir P. Lipp}
\altaffiliation{Current affiliation: Center for Free-Electron Laser Science CFEL, Deutsches Elektronen-Synchrotron DESY, 22607 Hamburg, Germany}
\author{Tobias Zier}
\author{Eeuwe S. Zijlstra}
\author{Martin E. Garcia}
\affiliation{Theoretical Physics and Center for Interdisciplinary Nanostructure Science and Technology (CINSaT), University of Kassel, Heinrich-Plett-Strasse 40, 34132 Kassel, Germany}


\date{\today}

\begin{abstract}
We develop an electronic-temperature dependent interatomic potential $\Phi (T_\text{e})$ for unexcited and laser-excited silicon.
The potential is designed to reproduce {\it ab initio} molecular dynamics simulations by requiring force- and energy matching for each time step.
$\Phi (T_\text{e})$ has a simple and flexible analytical form, can describe all relevant interactions and is applicable for any kind of boundary conditions (bulk, thin films, clusters).
Its overall shape is automatically adjusted by a self-learning procedure, which finally finds the global minimum in the parameter space.
We show that $\Phi (T_\text{e})$ can reproduce all thermal and nonthermal features provided by {\it ab initio} simulations.
We apply the potential to simulate laser-excited Si nanoparticles and find critical damping of their breathing modes due to nonthermal melting.
\end{abstract}

\keywords{silicon, interatomic potential, ultrafast melting, nonthermal effects, Density Functional Theory, Molecular Dynamics}

\maketitle

Structural phase transitions and material properties in thermodynamical equilibrium can often be remarkably well understood in terms of effective interatomic potentials depending only on the ionic coordinates \cite{Tersoff1988b,Stillinger1985,Gupta1980}.
Electrons, although microscopically responsible for bonding, are not explicitly treated as degrees of freedom. Effective interatomic potentials can be formally obtained in the framework of the Born-Oppenheimer approximation, assuming that the electrons are in their ground state.
A new situation arises, however, when solids are excited by intense ultrashort processes like femtosecond (fs) laser pulses or high-energy-ion bombardment.
In these cases, electrons are typically heated to a high temperature $T_\text{e}$, of the order of $T_\text{e} \gtrsim$ 1 eV = 11605 K, while ions remain at first at their temperature before excitation.
Such a transient nonequilibrium state can have a lifetime of the order of picoseconds (ps) \cite{Vechten1979} and lead to a variety of ultrafast nonthermal phenomena, like, for example, bond hardening or softening \cite{Recoules2006,Grigoryan2014,Fritz2007}, ultrafast structural solid-solid and solid-liquid phase transitions \cite{Cavalleri2001,Sciaini2009,Buzzi2018}, phonon squeezing \cite{Johnson2009,Zijlstra2013} and excitation of coherent phonons \cite{Cheng1991,Hase2003}.
Since these nonthermal effects are driven by excited electrons, it is clear that they cannot be described in terms of ground-state effective interatomic potentials.
Moreover, the same applies to the atomistic simulation of nanometer-scale fs-laser material processing \cite{Courvoisier2009,Wang2010,Hoehm2013,Hoehm2015}, since the transient nonequilibrium state initially created by the laser pulse plays an important role in the further evolution of the material being processed.

On the other hand, nonthermal phenomena in laser excited materials can be accurately described by {\it ab initio} molecular dynamics (MD) simulations \cite{Silvestrelli1996,Zijlstra2013b}, in which the ions move on a potential energy surface (PES) determined by electrons at finite (usually high) $T_\text{e}$.
The drawback of those simulations is that they are restricted to rather small supercells containing at most 10$^3$ atoms and are therefore not suitable for calculations on experimental length scales.
A possibility to extend the "range of action" of {\it ab initio} methods for a proper description of materials in the presence of hot electrons can be the derivation of an interatomic potential whose functional form depends on the degree of electronic excitation or, in the simplest case, on $T_\text{e}$.
Such an interatomic potential should correctly describe, apart from the structural properties of the laser-excited material, the evolution of bulk and surface after excitation at low computational cost, which makes them suitable for being used in large- and ultralarge scale MD simulations.
In spite of intensive research in this direction \cite{Khakshouri2008,Shokeen2011,Moriarty2012,Norman2012,Murphy2015,Darkins2018}, interatomic potentials fulfilling the above mentioned requirements could so far not be reliably constructed, partly due to the lack of sufficient microscopic data.

In this Letter, we develop, for the first time, an interatomic potential $\Phi (T_\text{e})$ that reliably describes Si at high $T_\text{e}$'s and that meets all above mentioned requirements.
$\Phi (T_\text{e})$ has a very simple analytical form, which automatically adjusts by a self learning procedure in order to reproduce the results of {\it ab initio} MD simulations with minimal error and computational effort.
As a first application of $\Phi (T_\text{e})$, we predict, by means of atomistic simulations, a damped breathing mode in laser-excited spherical Si nanoparticles.

To obtain sufficient {\it ab initio} data of the relevant atomic pathways for deriving $\Phi (T_\text{e})$, we performed many series of MD simulations using our in-house density-functional-theory (DFT) code CHIVES \cite{Grigoryan2014,Zijlstra2013b,Zijlstra2013}.
The DFT-MD simulations were performed on 320 Si atoms forming a thin film of 5.3-nm thickness. The thin-film geometry was achieved by using periodic boundary conditions and including vacuum in $z$-direction.
The supercell was prepared (thermalized) at an initial $T_\text{e}=T_\text{i} = 300$ K, being $T_\text{i}$ the ionic temperature \cite{supp-info}.
Then, starting from one randomly chosen initial condition taken from the thermalization run, 
MD simulations of the laser excited and unexcited dynamics were performed. In each run $T_\text{e}$ was kept constant. The time step was $2$ fs and the duration of the runs was 1 ps.
Eleven different $T_\text{e}$'s were considered in the range $316$ K ($1$ mHa) - $31577$ K ($100$ mHa).

The construction of a reliable potential for large-scale simulations needs a very good sampling of the available phase space.
Since the laser excitation of solids can lead to high local positive or negative pressures as well as local high/low densities (small/high interatomic distances) and since those extreme effects might not be captured by {\it ab initio} simulations on a small cell with homogeneous $T_\text{e}$, we performed the following additional {\it ab initio} MD runs to expand the sampling:
(i) the thin film at $T_\text{i}=300$ K was compressed stepwise and for the obtained structures the corresponding forces/energies were calculated at all eleven $T_\text{e}$'s; 
(ii) Since the film only expands significantly in MD simulations for $T_\text{e}$'s above the nonthermal melting threshold ($17052$ K \cite{Zier2014}), the coordinates from the MD simulation at $T_\text{e}=25262$ K were taken and used to calculate the forces/energies at all other $T_\text{e} < 25262$ K.
In total, around $10^6$ data-points were obtained for $T_\text{e}<25262$ K and $5\, 10^5$ for $T_\text{e}\geq 25262$ K.

\begin{figure}[htbp]
	\includegraphics[width=0.67\textwidth]{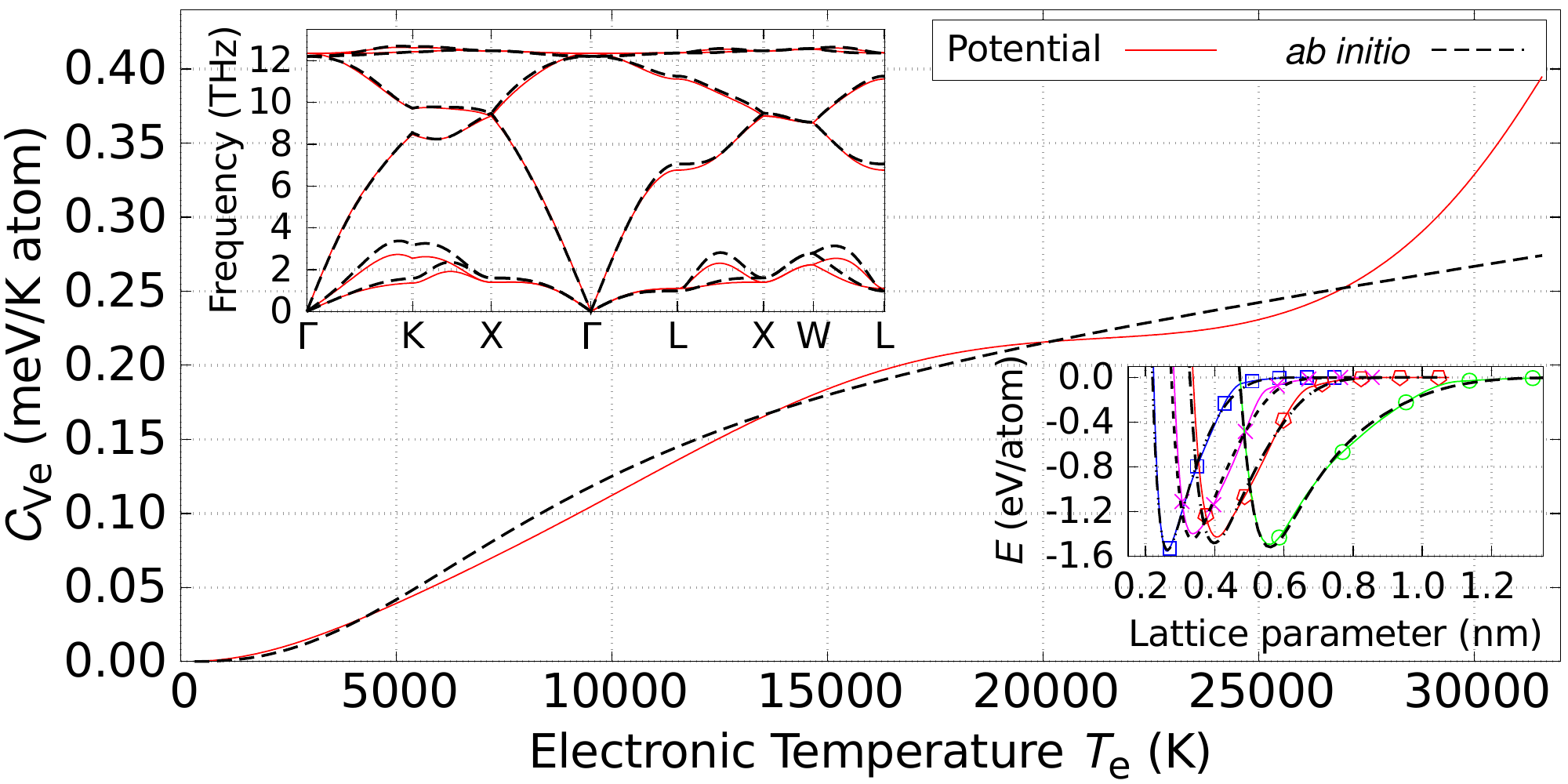}
	\caption{\label{fig:CVe_bs_Ecoh}
	Specific heat of the electrons $C_\text{Ve}$ as a function of $T_\text{e}$ for bulk Si.
	The upper inset indicates the phonon band structure at $T_\text{e}=18315$ K.
	The lower inset shows the Helmholtz free cohesive energy $E$ of the diamond (red with diamonds), fcc (magenta with crosses), bcc (green with circles) and sc (blue with squares) structures as a function of the lattice parameter at $T_\text{e}=18315$ K.
	Black dashed curves represent {\it ab initio} data and colored solid curves represent the values obtained from $\Phi(T_\text{e})$.}
\end{figure}

To obtain a reliable and physically appealing analytical form of $\Phi(T_\text{e})$, we construct it as a sum of different local interaction terms describing both covalent and metallic bonding, both present in Si.
We use two- and three-body interaction terms ($\Phi_2$ and $\Phi_3$, respectively) to describe covalent bonding, which is responsible for tetragonal bonding geometry due to $\mbox{sp}^3$ hybridization in ground state Si and was already very successfully described by these terms \cite{Stillinger1985}.
In addition, we use an embedding function $\Phi_\rho$, that calculates the local potential energy at an atomic site as a function of the surrounding atomic density $\rho$, and accounts for metallic bonding, which dominates in Si under pressure \cite{Wentorf1963} or in the molten phase \cite{Silvestrelli1996} and was already successfully described in metals by this function and a two-body interaction term \cite{Baskes1983,Zhakhovskii2009}.
We also add the Helmholtz free energy $\Phi_0$ of an isolated atom approximated directly from DFT.
$\Phi(T_\text{e})$ reads
\begin{eqnarray}
\Phi &=& \sum_{\scriptsize \begin{array}{c}i<j\\ r_{ij} < r^{(\text{c})}_2 \end{array}} \hspace{-10pt} \Phi_2(r_{ij}) + \sum_{\scriptsize \begin{array}{c}i\,j\,k\\ r_{ij},r_{ik} < r^{(\text{c})}_3 \end{array}} \hspace{-19pt} ' \ \Phi_3(r_{ij},r_{ik},\theta_{ijk}) \nonumber \\
 & & + \sum_{i} \Phi_{\rho}\left(\rho^{(2)}_i,\rho^{(3)}_i,\ldots,\rho^{\left(N^{(r)}_\rho\right)}_i\right) + \sum_{i} \Phi_0.
\label{equ:Phi}
\end{eqnarray}
Here $r_{ij}$ denotes the distance between atoms $i$ and $j$, $\theta_{ijk}$ is the angle between $\mathbf{r}_{ij}$ and $\mathbf{r}_{ik}$, the prime indicates that all summation indices are distinct, and $\rho^{(2)}_i$, $\rho^{(3)}_i, \ldots$ are different measures for the atomic density surrounding atom $i$ (see below).
Since the {\it ab initio} calculations yield that atomic interactions at large distance become negligible (see bottom inset of Fig. \ref{fig:CVe_bs_Ecoh}), we use for $\Phi_2$, $\Phi_3$ and $\Phi_\rho$ the individual cutoff radii $r^{(\text{c})}_2$, $r^{(\text{c})}_3$, $r^{(\text{c})}_\rho$, that are the distances beyond which the interaction between atoms is set to zero.

To keep $\Phi(T_\text{e})$ as simple and flexible as possible, we expand the terms $\Phi_2$, $\Phi_3$, $\Phi_\rho$ and also $\rho^{(2)}_i, \rho^{(3)}_i, \ldots$ into {\it polynomials}, which can, in principle, reproduce any physically reasonable function.
To achieve numerical stability, the polynomials must be functions of variables lying in the interval $[-1,1]$.
Hence, we use $\cos(\theta)$, where $\theta$ is a bond angle, and $1-r/r^{(\text{c})}$, being $r$ an interatomic distance, as variables for the polynomials.
The powers of the latter start from degree two to let $\Phi(T_\text{e})$ and its first derivatives continuously decreasing to zero as distances reach the cutoff radii.
Thus, $\Phi_2$, $ \Phi_3$ are constructed as
\begin{eqnarray}
	\Phi_2 &=& \sum_{q=2}^{N_2^{(r)}} c^{(q)}_2\, \left(1-\frac{r_{ij}}{r^{(\text{c)}}_2}\right)^q, 
	\label{equ:Phi_2} \\
	\Phi_3 &=& \sum_{q_1=2}^{N_3^{(r)}} \ \sum_{q_2=q_1}^{N_3^{(r)}} \ \sum_{q_3=0}^{N_3^{(\theta)}} c^{(q_1\,q_2\,q_3)}_{3} \, \times \nonumber \\
	& & \times\left(1-\frac{r_{ij}}{r^{(\text{c})}_3}\right)^{q_1} \left(1-\frac{r_{ik}}{r^{(\text{c})}_3}\right)^{q_2} \cos(\theta_{ijk})^{q_3},
	\label{equ:Phi_3}
\end{eqnarray}
and, for $q_1=2,3,\ldots, N^{(r)}_\rho$, the measures for the atomic density surrounding atom $i$ are constructed as
\begin{equation}
	\rho^{(q_1)}_i = \sum_{\scriptsize \begin{array}{c}j\neq i\\ r_{ij} < r^{(\text{c})}_\rho \end{array}} \left(1-\frac{r_{ij}}{r^{(\text{c})}_\rho}\right)^{q_1}.
	\label{equ:rho}
\end{equation}
By definition, $\rho^{(q_1)}_i \in [0,\infty)$, and if they are zero $\forall$, then $\Phi_\rho$ should be also zero.
Hence, to construct $\Phi_\rho$, we use powers of $\rho / (1+\rho)$, starting from degree one, for expanding the atomic density $\rho$, since $\rho / (1+\rho)$ is zero for $\rho=0$ and converges to one for $\rho\to \infty$.
\begin{equation}
	\Phi_\rho = \sum_{q_1=2}^{N^{(r)}_\rho} \ \sum_{q_2=1}^{N^{(\rho)}_\rho} c^{(q_1\,q_2)}_{\rho} \, \left(\frac{\rho^{(q_1)}_i}{1+\rho^{(q_1)}_i}\right)^{q_2}.
	\label{equ:Phi_rho}
\end{equation}

To determine the accuracy of $\Phi(T_\text{e})$ in reproducing the previously prepared {\it ab initio} runs $\{s\}$ at a given $T_\text{e}$, we define the following error function $W$ ({\it cf.} \cite{Ercole1994}), which is the weighted sum of the mean square relative errors in Helmholtz free cohesive energies and atomic forces of the different {\it ab initio} runs $\{s\}$ with weights $w^{(s)}_\text{E}$, $w^{(s)}_\text{f}$, respectively, obeying the sum rule $\sum_s w^{(s)}_\text{E}+\sum_s w^{(s)}_\text{f}=1$:
\begin{eqnarray}
	W &=& \sum_s w^{(s)}_\text{E}\, \frac{\sum \limits_{t} \Bigl(\Phi\bigl(\bigl\{\mathbf{r}^{(s)}_j(t)\bigr\}\bigr)- E^{(s)}(t)- \sum \limits_i \Phi_0 \Bigr)^2}{\sum \limits_{t} \left(E^{(s)}(t)\right)^2} \nonumber \\
	&& + \sum_s w^{(s)}_\text{f}\,\frac{\sum \limits_{t} \sum \limits_{i} \Bigl|-\nabla_{\mathbf{r}^{(s)}_i} \Phi\bigl(\bigl\{\mathbf{r}^{(s)}_j(t)\bigr\}\bigr) -\mathbf{f}^{(s)}_i(t)\Bigr|^2}{\sum \limits_{t} \sum \limits_{i} \bigl|\mathbf{f}^{(s)}_i(t)\bigr|^2}. \hspace{0.4cm}
	\label{equ:W}
\end{eqnarray}
Here, $\mathbf{r}^{(s)}_i$ denotes the position of atom $i$, $E^{(s)}(t)$ the total {\it ab initio} Helmholtz free cohesive energy of the system, and $\mathbf{f}^{(s)}_i(t)$ the {\it ab initio} force acting on atom $i$.

\begin{figure}[htbp]
	\includegraphics[width=0.67\textwidth]{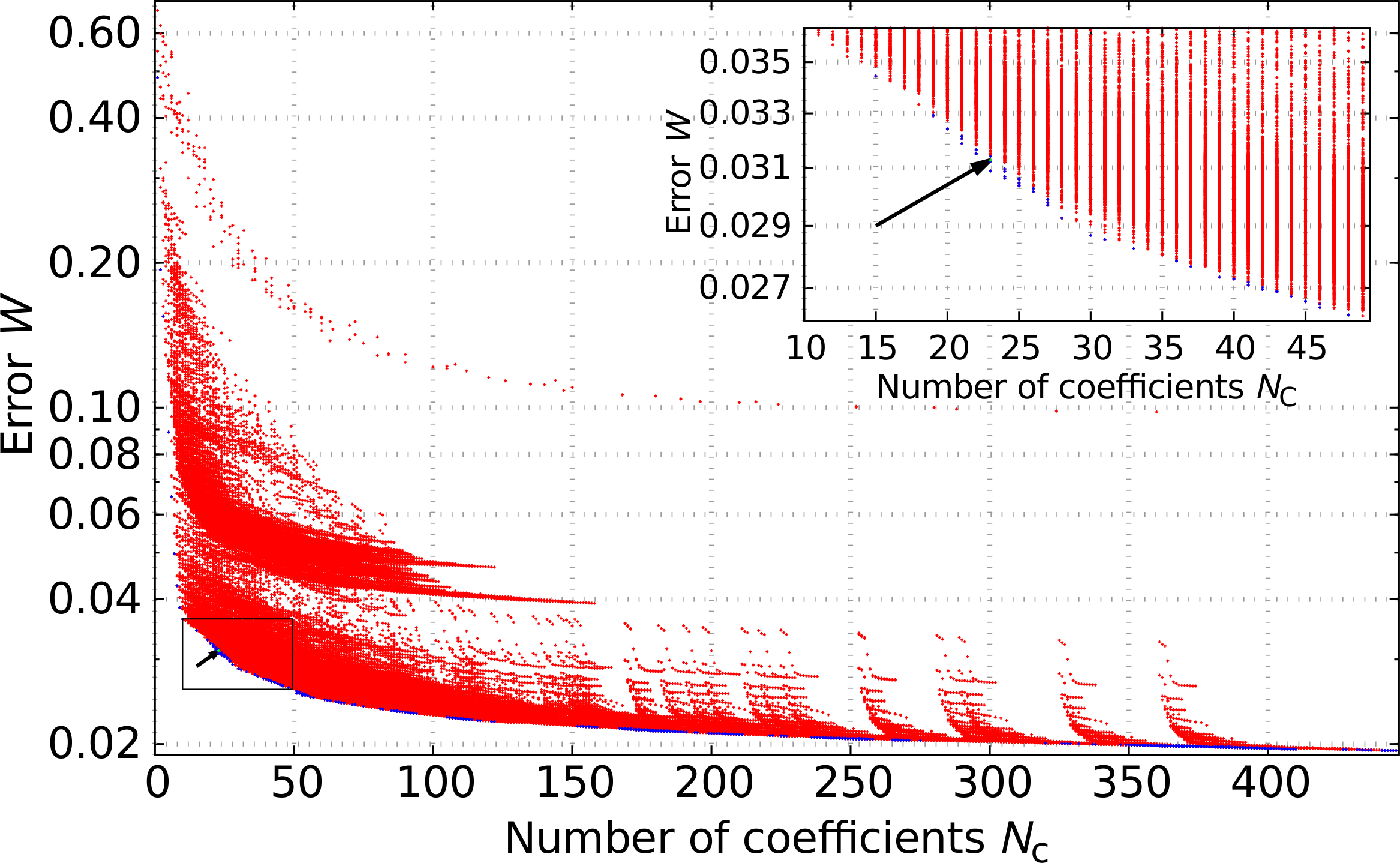}
	\caption{\label{fig:Werr}
	Error $W$ averaged over $T_\text{e}=316$ K and $T_\text{e}=18946$ K as a function of $N_\text{c}$ for all physically reasonable degree combinations. Each red dot represents an interatomic potential. The subset of potentials minimizing error and number of coefficients is highlighted in blue and the final choice, $\Phi(T_\text{e})$, is highlighted in green and marked by an arrow.}
\end{figure}

The analytical form of $\Phi(T_\text{e})$ is controlled by the polynomial degrees $N_2^{(r)}$, $N_3^{(r)}$, $N_3^{(\theta)}$, $N^{(r)}_\rho$, $N^{(\rho)}_\rho$.
Since these exhibit limits beyond which $W$ does not decreases significantly if degrees are further increased (see Fig. \ref{fig:Werr}), only a finite number of physical reasonable degree combinations exists.
Moreover, for any degree combination, the minimal $W$ and the corresponding optimal coefficients $\bigl\{c^{(q)}_2\bigr\}$, $\bigl\{c^{(q_1\,q_2\,q_3)}_3\bigr\}$, $\bigl\{c^{(q_1\,q_2)}_\rho\bigr\}$ and optimal cutoff radii can be derived.
This is possible on the basis of two facts:
Firstly, the optimal cutoff radii can be surely determined by a brute-force search, since physically reasonable cutoff radii lie in a finite interval and can be treated discretely because of the continuous dependence $W$ on them \footnote{$\Phi(T_\text{e})$ and the forces continuously decrease to zero as distances reach the cutoff radii}.
Secondly, for given cutoff radii, the associated optimal coefficients that minimize $W$ can always be uniquely found by solving a system of linear equations, since all coefficients are independent and appear linearly in $\Phi(T_\text{e})$ (see Eqs. 2-5).

The final choice of the polynomial degrees should lead to minimal $W$ and minimal computational cost.
Since many of the previous mentioned physical reasonable degree combinations do not fulfill these criteria, we select the final degrees from a subset containing only efficient degree combinations that exhibit a small $W$ and also a relatively small number of coefficients $N_\text{c}$.
This subset is automatically created by an iterative procedure from the set of all physically reasonable degree combinations. In our case, the total set of generated potentials amounts 165344, corresponding to all possible polynomial degree combinations.
The optimization (self-learning) procedure is initialized with the constant potential $\Phi=\sum_i \Phi_0$ with $N_\text{c} = 0$ and $W = 1$, and the degree combinations that maximize the error reduction per number of added coefficients $\Delta W/\Delta N_\text{c}$ are iteratively selected \cite{supp-info}.
Since the optimal degree combination of the polynomials should work well for all considered $T_\text{e}$'s, during the self-learning procedure we do not directly use Eq. (\ref{equ:W}) but rather the average of $W$ over $T_\text{e}=316$ K and $T_\text{e}=18946$ K.
Having obtained the subset of potentials exhibiting small error and reasonably small number of coefficients, we now chose the final polynomial degrees by manually checking to which extent the physical properties obtained in the {\it ab initio} calculations/simulations are reproduced by the different selected potentials at all considered $T_\text{e}$'s \cite{supp-info}.
Finally, we found the potential with the best performance, $\Phi(T_\text{e})$, which contains $N_\text{c}=23$ coefficients.

Since the resulting optimal cutoff radii vary insignificantly around $r^{(2)}_{c}=0.63$ nm, $r^{(3)}_{c}=0.42$ nm, $r^{(\rho)}_{c}=0.48$ nm at the eleven studied $T_\text{e}$'s, we chose these values and kept them constant for all $T_\text{e}$'s.
Thus, as a result of the optimization procedure, only the optimal coefficients depend on $T_\text{e}$.
These coefficients, which are only known at the eleven $T_\text{e}$'s, were approximated by polynomials of degree 5 in $T_\text{e}$ to obtain a continuous dependence on $T_\text{e}$ and be able to calculate the internal energy $U_\text{e}$ and the specific heat $C_\text{Ve}$ of the electrons per atom using the thermodynamic relations
\begin{equation}
	U_\text{e}=\Phi -T_\text{e}\,\frac{\partial \Phi}{\partial T_\text{e}}, \qquad C_\text{Ve} = -\frac{T_\text{e}}{N_\text{at}} \frac{\partial^2 \Phi}{\partial T^2_\text{e}},
	\label{equ:Ue_CVe}
\end{equation}
whereas $N_\text{at}$ is the total number of atoms.

Indeed, for the resulting final $\Phi(T_\text{e})$, which is provided as a table and a Fortran subroutine in the supporting information, both $U_\text{e}$ and $C_\text{Ve}$ are in very good agreement with the corresponding \textit{ab initio} results for bulk Si (see Fig. \ref{fig:CVe_bs_Ecoh} for $C_\text{Ve}$ and supporting information for $U_\text{e}$).
$\Phi(T_\text{e})$ also reproduces the phonon band structure of the bulk material as well as the cohesive energy curves of the diamond, fcc, bcc and sc structures for all studied $T_\text{e}$'s (see examples in the inset of Fig. \ref{fig:CVe_bs_Ecoh}). It is remarkable, that $\Phi(T_\text{e})$ does not only reproduce lattice properties but also electronic properties.

\begin{figure}[htbp] 
	\includegraphics[width=0.67\textwidth]{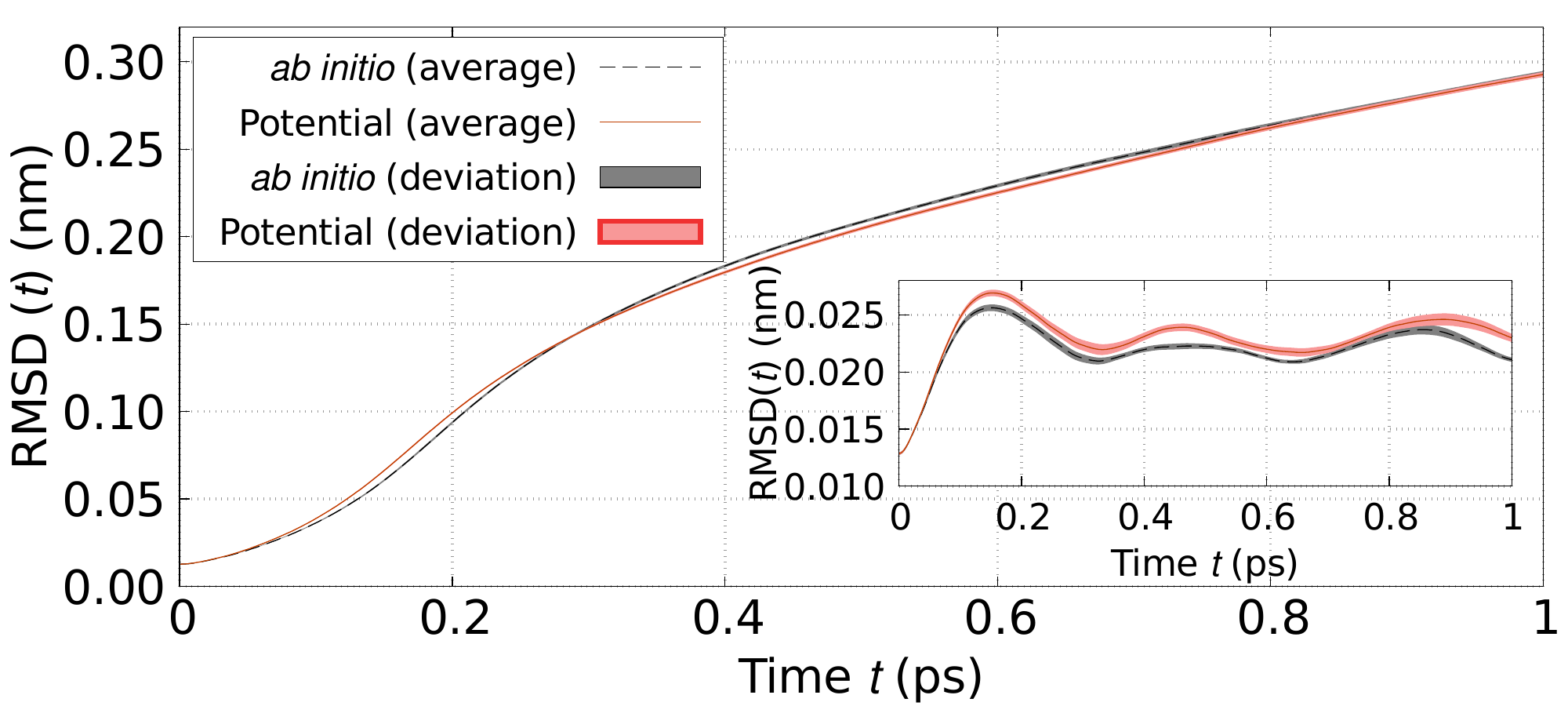}
	\caption{\label{fig:rmsd}
	Atomic root-mean-square displacements (RMSD) for bulk Si at $T_\text{e}=22104$ K from MD simulations performed {\it ab initio} (black dashed curve) and using $\Phi(T_\text{e})$ (red solid line) averaged over 40 runs.
	The inset shows the RMSD at $T_\text{e}=15789$ K averaged over 10 runs.}
\end{figure}

The relative error of $\Phi(T_\text{e})$ in the forces (second term of Eq. \eqref{equ:W}) decreases from 26\% to 6\% when $T_\text{e}$'s increases from $316$ K to $31577$ K.
This indicates, that the PES becomes less complex for higher $T_\text{e}$'s.
It is important to stress here that $\Phi(T_\text{e})$ describes forces in independent {\it ab initio} MD simulations, which were not used for its development, with the same accuracy.
The relative error in the Helmholtz free cohesive energies (first term of Eq. \eqref{equ:W}) lies always below 1.7\%.
The time evolution of the atomic root-mean-square displacements (RMSD) during laser-induced nonthermal melting \cite{Zijlstra2013b} and thermal phonon squeezing \cite{Zijlstra2013} is well described by $\Phi(T_\text{e})$ compared to {\it ab initio}, as it can be exemplary seen in Fig. \ref{fig:rmsd}.
Furthermore, the atomic RMSD perpendicular to the surface of the thin film after laser excitation is well reproduced by $\Phi(T_\text{e})$ at all $T_\text{e}$'s, indicating that $\Phi(T_\text{e})$ "knows" about the presence of a surface \cite{supp-info}.
$\Phi(T_\text{e})$ yields a very good description of elastic constants, pair-correlation function and bond angle distribution \cite{supp-info}.

By performing liquid crystal coexistence MD simulations for 65536 Si atoms with $\Phi(T_\text{e})$, we determined the melting temperature $T_\text{m}(p)= 1199 \pm 2\, \text{K} - 40\pm 3\, \frac{\text{K}}{\text{GPa}} \times p$ near zero pressure $p$.
Since $\Phi(T_\text{e})$ is developed from \textit{ab initio} simulations, $T_\text{m}$ differs from the experimental value, $T_\text{m}= 1687 \pm 5\, \text{K} - 58\, \frac{\text{K}}{\text{GPa}} \times p$ \cite{Yamaguchi2002,Jayaraman1963}. However, and thanks to the simple physical analytical form of $\Phi(T_\text{e})$, we can adjust the coefficients to reproduce the experimental $T_\text{m}$ without significant influences on the other properties \footnote{This adjustment is out of the scope of the present letter and will be content of a further publication.}.
Note, that such adjustment would be impossible to apply to the recently developed machine learning potentials \cite{Rupp2015,Behler2016,Grisafi2018}, which do not use a physical motivated analytical form.

Notice also, that previously developed potentials for Si at high $T_\text{e}$'s \cite{Shokeen2010,Shokeen2011,Darkins2018} exhibit a very inaccurate description of the atomic RMSD during thermal phonon squeezing and nonthermal melting in bulk Si and of the atomic RMSD perpendicular to the surface of the thin film \cite{Bauerhenne2018}.
Fitting the coefficients of the hereby used \cite{Kumagai2007} and of several widely used \cite{Stillinger1985,Tersoff1988b} classical analytical potentials to our thin-film DFT-MD simulations lead to a better description of the latter for the resulting potentials, but the accuracy of our $\Phi(T_\text{e})$ is not reached \cite{Bauerhenne2018}.

Having demonstrated the reliability of $\Phi(T_\text{e})$, we utilized it for classical MD simulations of laser-excited Si spherical nanoparticles on spatial scales which are typically not achievable with {\it ab initio} methods.
A Si nanosphere with the radius of $\sim 4.8$ nm and consisting of 23976 atoms was thermalized at $T_\text{e}=T_\text{i}=300$ K and zero pressure.
Then, we modeled two different laser excitations by instantly increasing $T_\text{e}$ from $300$ K to $13000$ K and $19000$ K. The use of a constant $T_\text{e}$ during the MD simulations after laser excitation can be physically justified in the context of ultrashort laser excitation: 
After being excited, electrons in Si quickly thermalize to a Fermi distribution \cite{Goldman1994}, and the subsequent incoherent electron-phonon coupling is expected to take a longer time, from 2 ps upwards depending on the number of excited electron-hole pairs \cite{Harb2006,Zijlstra2013}.
Using $U_\text{e}$ and $C_\text{Ve}$ from eq. \eqref{equ:Ue_CVe}, it would be possible to simulate the incoherent electron-phonon coupling in the context of the two temperature model \cite{Anisimov1974} with the only knowledge of $\Phi(T_\text{e})$. However, there is so far no clearly determined electron phonon-coupling constant $G_\text{ep}$ for Si.

\begin{figure}[htbp]
	\includegraphics[width=0.67\textwidth]{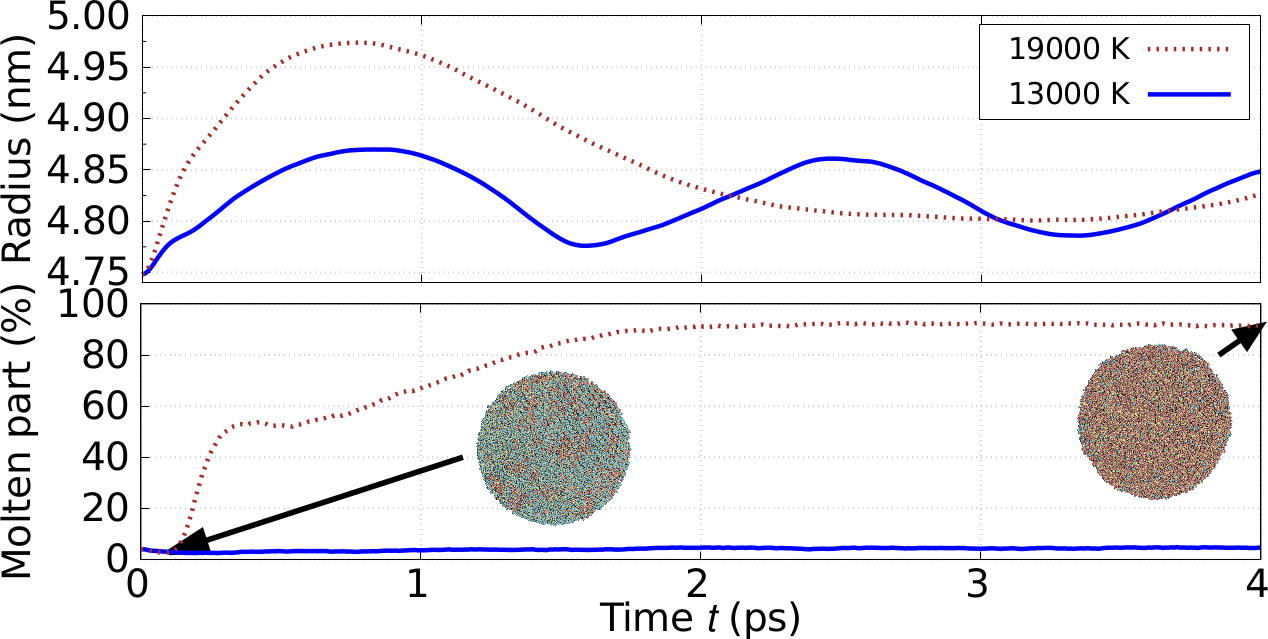}
	\caption{\label{fig:breathing_rmsd}
	Time evolution of the sphere radius (top) and percentage of molten material (bottom) in Si nanospheres at different $T_\text{e}$'s obtained from $\Phi(T_\text{e})$.
	In addition, the structure snapshots of the nanospheres are shown for 100 fs and 4 ps after the laser excitation in the case of $T_\text{e}=19000$ K.
	The atoms are colored by the csp value: blue corresponds to crystal environment, red to molten environment.}
\end{figure}

Fig. \ref{fig:breathing_rmsd} (top) shows the time evolution of the nanosphere radius at two $T_\text{e}$'s, showing the breathing modes in Si nanospheres upon laser excitation. 
Clearly, the amplitude of vibrations decreases for increasing time, being the damping very strong for high $T_\text{e}$. 
To explain this behavior, we investigated the structural state of the nanoparticles by using the central symmetry parameter (csp) \cite{Lipp2014}, which allows to distinguish whether a certain atom is surrounded by a crystalline (for $\text{csp}>0.968$) or by a liquid environment (for $\text{csp}<0.968$). 
csp allows to obtain the total percentage of molten material in the sphere depending on time for different $T_\text{e}$'s, which is shown in Fig. \ref{fig:breathing_rmsd} (bottom). The comparison of the two plots in Fig. \ref{fig:breathing_rmsd} suggests that the nonthermal melting strongly damps the laser-excited breathing modes in Si. 
Assuming that after 2ps incoherent electron-phonon heating can become important, and prevent further oscillations, our results suggest a sort of critical damping for $T_\text{e}=19000$ K.

In summary, we present in this paper an interatomic potential $\Phi(T_\text{e})$ for Si, which can describe the material in a wide range of $T_\text{e}$'s, including electrons in the ground-state and in laser excited states.
$\Phi(T_\text{e})$ utilizes a simple, physical motivated, and flexible analytical form and was developed auto-adjusted from DFT (LDA) simulations of the evolution of a thin film at various $T_\text{e}$'s.
$\Phi(T_\text{e})$ is able to accurately describe laser-driven effects caused by bond-softening, including nonthermal melting and thermal phonon squeezing.
In addition, it reliably reproduces cohesive energy curves for several bulk structures, phonon band structure, and elastic constants.
With the help of $\Phi(T_\text{e})$ we predict that nonthermal melting of Si nanospheres is accompanied by a strong damping of the breathing modes.
$\Phi(T_\text{e})$ can be use to describe laser processing of Si in the framework of the TTM-MD method \cite{Ivanov2003}.

\begin{acknowledgments}
The authors acknowledge the contribution by Dmitry S. Ivanov to the computer code utilized for classical MD simulations.
This work was supported by the DFG through the grant GA 465/18-1.
B.B. acknowledges the support by the "Promotionsstipendium des Otto-Braun Fonds" and by the "Abschlussstipendium der Universit\"at Kassel".
Computations were performed on the Lichtenberg High Performance Computer (HHLR) TU Darmstadt, on the IT Servicecenter (ITS) University of Kassel, and on the computing cluster FUCHS University of Frankfurt.
\end{acknowledgments}


%

\end{document}